\documentclass[11pt]{article}
\usepackage{amssymb}
\usepackage{amsmath}
\usepackage{amscd}
\usepackage{latexsym}

\catcode`@=11
\renewcommand{\section}{\@startsection{section}{1}{0pt}{\medskipamount}
{\medskipamount}{\large\bf}}
\numberwithin{equation}{section}
\catcode`@=12
\def\a{\alpha}
\def\b{\beta}
\def\D{\Delta}
\def\de{\delta}

\def\h{\eta}
\def\th{\theta}

\def\m{\mu}
\def\n{\nu}

\def\s{\sigma}
\def\p{\phi}

\def\La{\Lambda}
\def\ta{\tilde\alpha}
\def\tb{\tilde\beta}

\newcommand{\R}{\mathbb R}
\newcommand{\Hcal}{{\cal H}}

\def\ic{\mbox{i}}
\def\N2{$N{=}2$}
\def\pa{\mbox{$\partial$}}
\def\diff{\mbox{d}}

\def\sfrac#1#2{{\textstyle\frac{#1}{#2}}}
\def\>{\rangle}
\def\<{\langle}
\def\+{\dagger}
\def\={\ =\ }

\begin{document}

\begin{titlepage}
\setcounter{page}{0}

\vskip 2.0cm

\begin{center}

{\Large\bf Noncommutative Moduli for Multi-Instantons}

\vspace{15mm}

{\Large 
Tatiana A. Ivanova${}^\+$\,, \ 
Olaf Lechtenfeld${}^*$ \ and \ 
Helge M\"uller-Ebhardt${}^*$}
\\[10mm]
\noindent ${}^\+${\em Bogoliubov Laboratory of Theoretical Physics, JINR\\
141980 Dubna, Moscow Region, Russia}\\
{Email: ita@thsun1.jinr.ru}
\\[10mm]
\noindent ${}^*${\em Institut f\"ur Theoretische Physik,
Universit\"at Hannover \\
Appelstra\ss{}e 2, 30167 Hannover, Germany }\\
{Email: lechtenf, ebhardt@itp.uni-hannover.de}

\vspace{20mm}

\begin{abstract}
\noindent
There exists a recursive algorithm for constructing BPST-type multi-instantons
on commutative~$\R^4$. When deformed noncommutatively, however, 
it becomes difficult to write down non-singular instanton configurations 
with topological charge greater than one in explicit form. We circumvent 
this difficulty by allowing for the translational instanton moduli to become 
noncommutative as well. Such a scenario is natural in the self-dual Yang-Mills
hierarchy of integrable equations where the moduli of solutions are seen as
extended space-time coordinates associated with higher flows. By judicious 
adjustment of the moduli-noncommutativity we achieve the ADHM construction 
of generalized 't Hooft multi-instanton solutions with everywhere self-dual 
field strengths on noncommutative~$\R^4$.

\end{abstract}

\end{center}
\end{titlepage}

\section{Introduction} 

In recent years, many important nonperturbative field configurations,
like solitons, vortices, monopoles and instantons, have been generalized
in various dimensions to Moyal-type noncommutative spaces
(see e.g.~\cite{NS}--\cite{PSW} and reviews~\cite{DN} for further references).
Specializing to instantons on~$\R^4$, the self-dual Yang-Mills
equations~\cite{Belavin} are solved systematically through the ADHM
method~\cite{ADHM}. Its noncommutative extension to $\R^4_\th$, as developed
in~\cite{NS}--\cite{Furu} and \cite{Nekrasov}--\cite{TZ}, is straightforward
for self-dual~$\th$ but needs modification in case of anti-self-dual~$\th$.

A singular subclass of solutions are the 't~Hooft multi-instantons, which have
also been deformed noncommutatively by way of the splitting and ADHM
approaches~\cite{LP1,TZ}. With the help of Murray-von-Neumann transformations 
-- the noncommutative analog of singular gauge transformations --
one can remove the singularities and arrive at a non-singular gauge~\cite{LP1}.
A direct path to non-singular 't~Hooft multi-instantons is again offered
by the ADHM construction. Its noncommutative extension (for self-dual~$\th$)
yields a recursive algorithm for generating $n$-instanton configurations on 
$\R^4_\th$; yet, the explicit realization is rather technical beyond~$n{=}1$. 

In this letter we propose a generalization of noncommutative multi-instantons
by rendering part of the instanton {\it moduli noncommutative\/}. 
This is a logical step in the self-dual Yang-Mills hierarchy where moduli 
are naturally regarded as additional space-time coordinates. Surprisingly, 
with a noncommutative (translational) moduli space it is possible to write 
down explicit 't~Hooft $n$-instanton configurations with field strengths 
being self-dual everywhere, and we will do this here.

\medskip

\section{Commutative non-singular multi-instantons}

\noindent
{\bf Notation.\ }
Instantons are localized finite-action solutions to the classical
equations of motion for a Euclidean field theory \cite{Belavin,Raj}. 
In this paper we specialize to four-dimensional Euclidean Yang-Mills theory 
with the gauge group~U(2). Hence, we have four $u(2)$-valued gauge potentials 
$A_{\mu}$ and the field strengths
$\ F_{\mu\nu}=\pa_\mu A_\nu - \pa_\nu A_\mu + [A_\mu,A_\nu]$. 
Possible solutions to the field equations $D_{\mu}F_{\mu\nu}=0$ are obtained 
by demanding the field strength to be self-dual,
\begin{equation} \label{sd} 
\widetilde F_{\mu \nu}\ \equiv\ \sfrac{1}{2} \epsilon_{\mu\nu\lambda\rho}
F_{\lambda\rho} \= F_{\mu\nu} \qquad\textrm{for}\quad \mu,\nu,\ldots=1,2,3,4\ ,
\end{equation} 
since for such fields the equations of motion will be satisfied
due to the Bianchi identities $\ D_{\mu}\widetilde F_{\mu\nu}=0.$ 

It is convenient to introduce a few abbreviations in order to simplify the
expressions for the explicit solutions to~(\ref{sd}) we are going to recall 
momentarily.
Let us introduce the matrices
\begin{equation}\label{ee+}
\bigl(e_\mu\bigr)\ =\ \bigl(-\ic\s_a\,,\,{\bf1}_2\bigr)
\qquad\textrm{and}\qquad
\bigl(e^\+_\mu\bigr)\ =\ \bigl(\ic\s_a\,,\,{\bf1}_2\bigr)
\end{equation}
which enjoy the properties
\begin{equation}\label{prop_ee}
\begin{aligned}
e^\+_\mu\,e_\nu\ &=\ \de_{\mu\nu}{\bf1}_2 + \h^a_{\mu\nu}\,\ic\s_a\
=:\ \de_{\mu\nu}{\bf1}_2 + \h_{\mu\nu} \ ,\\ 
e_\mu\,e^\+_\nu\ &=\ \de_{\mu\nu}{\bf1}_2 + \bar{\h}^a_{\mu\nu}\,\ic\s_a\
=:\ \de_{\mu\nu}{\bf1}_2 + \bar{\h}_{\mu\nu} \ ,
\end{aligned}
\end{equation}
where $\s_a$, $a=1,2,3$, are the Pauli matrices while
$\h^a_{\mu\nu}$ and $\bar\eta^a_{\mu\nu}$ denote the
self-dual and anti-self-dual 't~Hooft tensors~\cite{Prasad}, 
respectively, which satisfy the identities
\begin{equation}\label{prop_h}
\h^a_{\m\n}\bar\h^b_{\m\n}\=0\qquad\textrm{and}\qquad 
\h^a_{\m\n}\h^b_{\m\n}\=4\de^{ab}\ .
\end{equation}
With the help of the matrices~(\ref{ee+}), one forms the quaternions
\begin{equation}
x\ :=\ x^\mu e^\+_\mu \qquad\textrm{and}\qquad
x^\+ \= x^\mu e_\mu \qquad\textrm{with}\quad \{x^\mu\}\in\R^4 \ .
\end{equation}
We will frequently have to shift $x^\mu$ by some real constants~$a_i^\mu$,
$i=1,\ldots,n$, and so define 
\begin{equation}
x^\mu_i\ :=\ x^\mu - a_i^\mu \qquad\Longrightarrow\qquad
x_i \= x_i^\mu e^\+_\mu \qquad\textrm{and}\qquad a_i \= a_i^\mu e^\+_\mu\ .
\end{equation}
Likewise, in addition to the radius-squared $r^2:=x^\mu x^\mu$ we introduce
the distance-squared to the point~$a_i$ as
\begin{equation}
r_i^2\ :=\ x_i^\mu x_i^\mu \= (x^\mu-a_i^\mu)(x^\mu-a_i^\mu)
\qquad\qquad\textrm{(no sum over $i$)} \ .
\end{equation}

\medskip

\noindent
{\bf ADHM construction.\ }
The most systematic way to generate instanton solutions is via 
the ADHM approach.
The construction (see~\cite{ADHM, Prasad})
of an $n$-instanton solution is based on a 
\begin{equation} \label{matrices}
(2n{+}2)\times2 \textrm{\ matrix\ \ } \Psi \qquad\textrm{and a}\quad
(2n{+}2)\times2n \textrm{\ matrix\ }\quad 
\Delta\={\bf a}+{\bf b}(x{\otimes}{\bf1}_n) \ ,
\end{equation}
where $\bf a$ and~$\bf b$ are constant $(2n{+}2)\times 2n$ matrices.
These matrices must satisfy the following conditions:
\begin{align}
\Delta^\+\Delta\quad& \textrm{is invertible}\ ,\label{c1}\\
[\,\Delta^\+\Delta\,,\,e_\mu\otimes{\bf1}_n\,]\ &=\ 0 \quad\forall x\ , 
\label{c2}\\
\Delta^\+\Psi\ &=\ 0\ ,\label{c3}\\
\Psi^\+\Psi\ &=\ {\bf1}_2 \ .\label{c4}
\end{align}
It is not difficult to see that conditions (\ref{c1}) and~(\ref{c2})
are met if
\begin{equation} \label{c5}
\Delta^\+\Delta\ =\ {\bf1}_2 \otimes h^{-1}_{n\times n}\ .
\end{equation}
For $(\Delta,\Psi)$ satisfying~(\ref{c1})--(\ref{c4})
the gauge potential is chosen in the form
\begin{equation} \label{adhmA}
A\ =\ \Psi^\+\,\diff\Psi\ .
\end{equation}
The resulting gauge field~$F$ will be self-dual if $\Delta$ and~$\Psi$
obey the completeness relation
\begin{equation} \label{complete}
\Psi\,\Psi^\+\ +\ \Delta\,(\Delta^\+\Delta)^{-1}\Delta^\+\ =\ {\bf1}_{2n+2}\ .
\end{equation}
Namely, using (\ref{c3}), (\ref{c4}) and (\ref{complete}), we find
\begin{equation} \label{adhmF}
F_{\mu\nu}\ =\
\pa_\mu(\Psi^\+\pa_\nu\Psi)\ -\ \pa_\nu(\Psi^\+\pa_\mu\Psi)\
+\ [\,\Psi^\+\pa_\mu\Psi\,,\,\Psi^\+\pa_\nu\Psi\,] 
\ =\ 2\,\Psi^\+ {\bf b}\,(\Delta^\+\Delta)^{-1} \h_{\mu\nu}\,{\bf b}^\+\Psi\ ,
\end{equation}
i.e. the anti-self-dual part of~$F_{\mu\nu}$ is zero.

To become more concrete, let us take the following ansatz 
(see e.g.~\cite{Corrigan,LP1}):
\begin{equation} \label{ansatz}
\Psi\ =\ \begin{pmatrix}
\Psi_0 \\ \Psi_1 \\ \vdots \\ \Psi_n \end{pmatrix}
\ ,\qquad
{\bf a}\ =\ \begin{pmatrix}
\La_1{\bf1}_2 & \ldots & \La_n{\bf1}_2 \\
-a_1          &        & {\bf0}_2      \\
              & \ddots &               \\
{\bf0}_2      &        & -a_n          \end{pmatrix}
\qquad\textrm{and}\qquad
{\bf b}\ =\ \begin{pmatrix}
{\bf0}_2 & \ldots & {\bf0}_2 \\
{\bf1}_2 &        & {\bf0}_2 \\
         & \ddots &          \\
{\bf0}_2 &        & {\bf1}_2 \end{pmatrix}
\end{equation}
where $a_i=a_i^\mu e_\mu^\+$ and the $\La_i$ are positive constants 
(scale parameters).
{}From (\ref{ansatz}) we get
\begin{equation} \label{deltaform}
{\D}\ =\ \begin{pmatrix}
\La_1{\bf1}_2 & \ldots & \La_n{\bf1}_2 \\
x_1           &        & {\bf0}_2      \\
              & \ddots &               \\
{\bf0}_2      &        & x_n          \end{pmatrix}
\qquad\textrm{and}\qquad
{\D^\+}\ =\ \begin{pmatrix}
\La_1{\bf1}_2 & x_1^\+   &        & {\bf0}_2 \\
   \vdots     &          & \ddots &          \\
\La_n{\bf1}_2 & {\bf0}_2 &        & x_n^\+   \end{pmatrix}\ ,
\end{equation}
and arrive at
\begin{equation} \label{c5res}
\Delta^\+\Delta\ =\ {\bf1}_2 \otimes (\de_{ij}r_j^2 + \La_i\La_j)\
=:\ {\bf1}_2 \otimes (R + \La\La^T)\ ,
\end{equation}
where
\begin{equation}
R\ =\ \begin{pmatrix}
r_1^2 &        & 0     \\
      & \ddots &       \\
0     &        & r_n^2 \end{pmatrix} 
\qquad\textrm{and}\qquad
\La\ =\ \begin{pmatrix}
\La_1 \\  \vdots \\ \La_n \end{pmatrix} \ .
\end{equation}
{}From (\ref{c5res}) we see that the condition (\ref{c5}) (and thus also
(\ref{c1}) and (\ref{c2})) is satisfied.
Indeed, by direct calculation one finds
\begin{equation} \label{h}
{\bf1}_2\otimes h_{n\times n}\ =\ (\Delta^\+\Delta)^{-1}\ =\
{\bf1}_2\otimes\bigl( R^{-1} - R^{-1}\La\,\phi_n^{-1}\La^T R^{-1} \bigr)
\end{equation}
with
\begin{equation} \label{phin}
\phi_n\ =\ 1+\ \sum_{i=1}^n \frac{\La_i^2}{r_i^2}\ .
\end{equation}

For the given form (\ref{deltaform}) of $\Delta$, 
the remaining conditions (\ref{c3}) and (\ref{c4}) become
\begin{align} \label{a}
&\La_i\Psi_0 + x_i^\+\Psi_i\={\bf0}_2 
\qquad\textrm{for}\quad i=1,\ldots,n\ ,\\ \label{b}
&\Psi_0^\+\Psi_0 + \Psi_1^\+\Psi_1 +\ldots+ \Psi_n^\+\Psi_n \= {\bf1}_{2}\ .
\end{align}
The task is to solve these two equations.
If successful one can evaluate the gauge potential~(\ref{adhmA})
and its field strength~(\ref{adhmF}). The latter is guaranteed to be
self-dual since the completeness relation (\ref{complete})
is automatically satisfied in the commutative case.

\medskip

\noindent
\textbf{One instanton.\ } 
One starts with the ansatz
\begin{equation} \label{oneansatz}
\Psi_0 \= x_1^\+ f_1 \qquad\textrm{and}\qquad \Psi_1 \= -\La_1 f_1
\end{equation}
which solves (\ref{a}) for an arbitrary matrix-valued function~$f_1$. 
With $\ x_1 x_1^\+ = r_1^2 {\bf1}_2$, 
the normalization (\ref{b}) then determines this function 
(up to a constant unitary matrix) as
\begin{equation}
f_1 \= \frac{1}{\sqrt{r_1^2 + \La_1^2}} \, {\bf1}_2 \ .
\end{equation}
This solution is obviously non-singular at finite values of $r_1^2$.
Using~(\ref{adhmA}) one arrives at 
\begin{equation}
A_{\mu} \= \Psi_0^\+\,\pa_\mu \Psi_0 + \Psi_1^\+\,\pa_\mu \Psi_1 
\= - \eta_{\mu\nu} \ \frac{x_1^{\nu}}{r_1^2 + \La_1^2} \ ,
\end{equation}
which is known as the BPST solution \cite{Belavin}.

\medskip

\noindent
\textbf{Two instantons.\ } 
In this case one takes the ansatz
\begin{equation} \label{twoansatz}
\Psi_0 \= x_1^\+ (a_2{-}a_1)\, x_2^\+ f_2 \ ,\quad 
\Psi_1 \= -\La_1 (a_2{-}a_1)\, x_2^\+ f_2 \quad\textrm{and}\quad 
\Psi_2 \= -\La_2 (a_2{-}a_1)\, x_1^\+ f_2\ ,
\end{equation}
which again solves (\ref{a}) for any matrix-valued function~$f_2$. With this, 
(\ref{b}) yields the function
\begin{equation}
f_2 \= \frac{1}{\sqrt{(a_2{-}a_1)^2 \,
(r_1^2\,r_2^2 + \La_1^2\,r_2^2 + \La_2^2\,r_1^2)}} \, {\bf1}_2 \ .
\end{equation}
This configuration too is non-singular since $r_1^2$ and $r_2^2$ cannot 
vanish simultaneously.

\medskip

\noindent
\textbf{Recursion for $n$ instantons.\ } 
It is possible to systematize the above sequence of ans\"atze and
write down a recursive formula for the $n$-instanton solution.
Because its explicit form is somewhat intricate and we will not make
use of it later on, there is no point displaying it here. It may be
remarked, however, that the proof of regularity (at finite points) is 
rather non-trivial. Nevertheless, it is possible in this way to generate 
explicit non-singular multi-instanton configurations for any instanton number.
To be sure, the singular 't Hooft configurations are also easily obtained
within the ADHM approach, with appropriate ans\"atze for~$\Psi_i$
(see, e.g.~\cite{Corrigan}).

\medskip

\section{Noncommutative non-singular multi-instantons}

In this section we shall give a short account of the status of
explicit noncommutative multi-instanton configurations for the 
(simpler) case of self-dual noncommutativity.

A Moyal deformation of Euclidean~$\R^4$ is achieved by replacing
the ordinary pointwise product of functions on it by the nonlocal
but associative Moyal star product. The latter is characterized by
a constant antisymmetric matrix $(\th^{\mu\nu})$ which prominently 
appears in the star commutation relation between the coordinates,
\begin{equation} \label{nccoord}
[\,x^\mu\,,\,x^\nu\,] \= \ic\,\th^{\mu\nu} \ .
\end{equation}
A different realization of this algebraic structure keeps the standard
product but promotes the coordinates (and thus all their functions)
to noncommuting operators acting in an auxiliary Fock space~$\Hcal$.
The two formulations are tightly connected through the Moyal-Weyl map.
When dealing with noncommutative U(2) Yang-Mills theory from now on,
we shall not denote the noncommutativity by either inserting stars in all
products or by putting hats on all operator-valued objects, but simply
by agreeing that our coordinates are subject to~(\ref{nccoord}).
The existence of~$(\th^{\mu\nu})$ breaks the Euclidean SO(4) symmetry
to SO(4) $\cap$ Sp(4,$\R$) $=$ U(2), but we may employ SO(4) rotations 
to pick a basis in which $(\th^{\mu\nu})$ takes Darboux form, i.e.~the only
nonzero entries are $\th^{12}=-\th^{21}$ and $\th^{34}=-\th^{43}$.
Such a matrix is a linear combination of the self-dual $(\h^{3\mu\nu})$ 
and the anti-self-dual $(\bar\h^{3\mu\nu})$.
In this work, we restrict ourselves to the special case of a purely 
self-dual noncommutativity tensor given by
\begin{equation}\label{theta}
\th^{\mu\nu}\ =\ \th\,\h^{3\mu\nu}\ .
\end{equation}

Let us try to generalize the ADHM construction of the previous section 
to the noncommutative case. We take the same multi-instanton ans\"atze
as in the commutative case but must take care of ordering now.
It is quickly verified that (for the the above choice of~$(\th^{\mu\nu})$)
one actually ends up with the same equations 
(\ref{matrices})--(\ref{adhmF}) and (\ref{a})--(\ref{b}),
of course now holding for noncommutative coordinates. In contrast to
the previous section, the completeness relation (\ref{complete}) is
no longer automatic, and so one needs to show that it holds as well.

\medskip

\noindent
\textbf{One instanton.\ } 
This was already calculated by Furuuchi \cite{Furu}. 
Irrespective of the noncommutativity, the ansatz~(\ref{oneansatz})
solves (\ref{a}) for any matrix $f_1$. 
The determination of~$f_1$ again proceeds via~(\ref{b})
but one must take into account a modified relation,
\begin{equation} \label{xdagx}
x_i^\+\,x_i \= r_i^2\,{\bf1}_2 \qquad\textrm{but}\qquad
x_i\,x_i^\+ \= r_i^2{\bf1}_2 - 2\th\,\sigma_3 
\qquad\qquad\textrm{(no sum over $i$)}\ ,
\end{equation}
since coordinate products now feature antisymmetric parts. The result is
\begin{equation} f_1 \= \Biggl( \begin{array}{cc}
\frac{1}{\sqrt{r_1^2 - 2 \theta + \Lambda_1^2}} & 0
\\ 0 & \frac{1}{\sqrt{r_1^2 + 2 \theta + \Lambda_1^2}}
\end{array} \Biggr) \ .
\end{equation}
This is non-singular because the spectrum of the operator $r_1^2$ is
bounded by~$2\th$. Furthermore,
\begin{equation}
\Psi \,\Psi^{\dagger} \,=\, \biggl( \begin{array}{cc} 
x_1^{\dagger} f_1^2 \, x_1 & -\Lambda_1 x_1^{\dagger} f_1^2 \\[6pt] 
-\Lambda_1 f_1^2 \, x_1 & \Lambda_1^2 f_1^2 \end{array} \biggr) 
\qquad\textrm{and}\qquad 
\Delta \, (\Delta^{\dagger} \Delta)^{-1} \Delta^{\dagger} \,=\, \biggl(
\begin{array}{cc} \frac{\Lambda_1^2}{r_1^2 + \Lambda_1^2}{\bf1}_2 &
\frac{\Lambda_1}{r_1^2 + \Lambda_1^2} x_1^{\dagger} \\[6pt] 
x_1 \frac{\Lambda_1}{r_1^2 + \Lambda_1^2} & 
x_1 \frac{1}{r_1^2 + \Lambda_1^2} x_1^{\dagger} \end{array} \biggr)
\end{equation}
so that, due to $\ f_1^2\,x_1 = x_1 \frac{1}{r_1^2 +\La_1^2}$,
the completeness relation (\ref{complete}) is indeed satisfied.
Finally one can calculate from (\ref{adhmA}) the gauge potential,
which is entirely regular and merges with the BPST solution for $\th\to0$.
The explicit expression coincides with the one obtained in the dressing 
approach by Horv\'ath et al in~\cite{FSI}. 

\medskip

\noindent
\textbf{Two instantons.\ } 
As in commutative case one takes the ansatz~(\ref{twoansatz})
which fulfils (\ref{a}) also when read noncommutatively. 
In order to find~$f_2$ one then computes
$\Psi_0^\+\Psi_0+\Psi_1^\+\Psi_1+\Psi_2^\+\Psi_2$ which yields
a non-diagonal $2{\times}2$ matrix with noncommuting matrix elements.
Equating it to unity and solving for~$f_2$ turns out to be technically
difficult, and we will not try to do this here. Nevertheless, 
we expect this solution to be non-singular and the completeness 
relation~(\ref{complete}) to hold for it.

\medskip

\noindent
\textbf{Multi-instantons.\ } 
The recursive construction mentioned in the previous section can be carried 
over to the noncommutative domain.  Yet, the determination of the 
matrices~$f_n$ and the verification of the completeness relation gets 
increasingly complicated due to the noncommutativity. 
To summarize this section, for the gauge group U(2) BPST-type instanton 
solutions on noncommutative~$\R^4$ are known only for charge~one. 
It remains a computational challenge to present an explicit noncommutative 
U(2) two-instanton configuration. 

\medskip

\section{Multi-instantons with noncommutative translational moduli}

\noindent
{\bf Noncommutative moduli.\ }
In this section we propose an unorthodox alternative which avoids all
previously mentioned difficulties. It employs the noncommutative 't Hooft
multi-instantons (via ADHM) but allows their translational moduli to
become noncommutative as well! The result is, of course, a rather
non-standard generalization of multi-instantons.

To be specific, we now modify the commutation relations to
\begin{equation}\label{acom}
[x^\m , x^\n ]= \ic\th^{\mu\nu}\ ,\qquad
[a_i^\m , a_j^\n ]= -\ic\th^{\mu\nu}\de_{ij} 
\qquad\textrm{and}\qquad [x^\m , a_j^\m ]=0\ ,
\end{equation}
where $\th^{\mu\nu}$ is given in (\ref{theta}). As a consequence,
\begin{equation}\label{prop_x}
[x_i^\m , x_j^\n ]\ =\ \ic\th^{\m\n}\,(1-\de_{ij}) 
\qquad\textrm{for}\quad i,j = 1,\ldots,n \ .
\end{equation}
Using (\ref{prop_ee}), (\ref{prop_h}) and (\ref{prop_x})
we obtain -- in distinction to (\ref{xdagx}) --
\begin{equation}\label{prod_x}
x_i^\+ x_i \= x_i\,x_i^\+ \= 
x_i^\m x_i^\m \,{\bf1}_2  \= r^2_i \,{\bf1}_2
\qquad\qquad\textrm{(no sum over $i$)} \ .
\end{equation}

\medskip

\noindent
{\bf Invertibility of $r^2_i$.\ } 
The commutation relations (\ref{acom}) can be 
realized in terms of annihilation and creation operators,
\begin{equation}\label{i1}
\{x^\m\}\ \mapsto\ \{\a_0,\b_0,\a_0^\+,\b_0^\+\} \qquad\textrm{and}\qquad 
\{a_i^\m\}\ \mapsto\ \{\a_i,\b_i,\a_i^\+,\b_i^\+\} \ ,
\end{equation}
acting in the tensor product 
$\Hcal :=\Hcal_0\otimes\Hcal_1\otimes\ldots\otimes\Hcal_n$
of $n{+}1$ copies of the two-oscillator Fock space. 
In this formulation one finds that
\begin{equation}\label{i2}
r_i^2\=2\th \bigl[(\a_i^\+ - \a_0)(\a_i - \a_0^\+) + 
(\b_i^\+ - \b_0)(\b_i - \b_0^\+)\bigr] \= 2\th\, (\ta^\+_i\ta_i + 
\tb^\+_i\tb_i)
\end{equation}
with new annihilation operators
\begin{equation}\label{i3}
\ta_i\ :=\ \a_i - \a_0^\+\qquad\mbox{and}\qquad\tb_i\ :=\ \b_i - \b_0^\+
\end{equation}
created by a Bogoliubov transformation. The general form 
of such a transformation is (see e.g.~\cite{Per})
\begin{equation}\label{i4}
\begin{pmatrix}\ta_0 \\ \ta_i \end{pmatrix}\=
\begin{pmatrix}\a_0 \\ \a_i \end{pmatrix}\, +\,
\begin{pmatrix}b_{11}&b_{12} \\ b_{12}&b_{22} \end{pmatrix}
\biggl(\begin{matrix}\a_0^\+ \\ \a_i^\+ \end{matrix}\biggr) 
\qquad\textrm{and}\qquad
\begin{pmatrix}\tb_0 \\ \tb_i \end{pmatrix}\=
\begin{pmatrix}\b_0 \\ \b_i \end{pmatrix}\, +\,
\begin{pmatrix}b_{11}&b_{12} \\ b_{12}&b_{22} \end{pmatrix}
\biggl(\begin{matrix}\b_0^\+ \\ \b_i^\+ \end{matrix}\biggr)
\end{equation}
where $\ (b_{ij})=:B\ $ is a symmetric $2\times 2$ matrix with complex 
entries. Note that the new annihilation operators $\ta_0$ and $\ta_i$ 
have normalizable vectors in their kernel only if the hermitian matrix 
$\ {\bf1}_2-BB^\+\ $ is positive definite. 
The case under consideration, however, is degenerate because
\begin{equation}\label{i5}
B\=\begin{pmatrix}0& -1 \\ -1&0 \end{pmatrix}
\qquad\Longrightarrow\qquad
{\bf1}_2 - BB^\+ \={\bf0}_2 \ .
\end{equation}
Hence, the operators $\ \a_i{-}\a_0^\+\ $ and $\ \b_i{-}\b_0^\+\ $ 
as well as their hermitian conjugates have no zero modes either on $\Hcal$ 
or on its dual $\Hcal^*$.
We conclude that, for any $i=1,\ldots,n$, the operator $r^2_i$ is invertible
on all finite-norm states, i.e.~on $\Hcal$ as well as on $\Hcal^*$.

\medskip

\noindent
{\bf Ansatz.\ } 
Since for self-dual $(\th^{\mu\nu})$ the ADHM scheme described in the previous
section is unaltered, we take over the ansatz~(\ref{ansatz}) unchanged,
in order to construct noncommutative 't~Hooft $n$-instantons
with noncommutative moduli parameters (\ref{acom}).
The commutative computation can be literally copied until arriving at
(\ref{h}) with (\ref{phin}), after having assured that $r_i^2$ is indeed
invertible in the noncommutative case.

The task to solve the equations (\ref{a}) and~(\ref{b}) is accomplished with
\begin{equation} \label{Psisol}
\Psi_0\ =\ \phi_n^{-\frac12}\,{\bf1}_2 \qquad\textrm{and}\qquad
\Psi_i\ =\ -x_i\,\frac{\La_i}{r_i^2}\,\phi_n^{-\frac12}\ ,
\end{equation}
where the factor $\phi_n^{-\frac12}$ was introduced 
to achieve the normalization
\begin{equation}
\Psi^\+\Psi\ =\ \phi_n^{-\frac12}\Bigl(
1+\ \sum_{i=1}^n\frac{\La_i^2}{r_i^2} \Bigr){\bf1}_2\;\phi_n^{-\frac12}\ 
=\ {\bf1}_2\ .
\end{equation}
Hence, our $(\Delta,\Psi)$ satisfies all conditions (\ref{c1})--(\ref{c5}),
and we can define the gauge potential via~(\ref{adhmA}). Note that this 
configuration with noncommutative translational moduli can be interpreted 
as a four-dimensional ``slice'' of a solution to the generalized self-dual 
Yang-Mills equations on $\R^{4+4n}$~\cite{Ward} in which both $x^\m$ 
and $a^\m_i$ play the role of (noncommutative) coordinates.

\medskip

\noindent
{\bf Completeness relation.\ } 
To be sure that we have indeed constructed self-dual field configurations, 
we must still check the completeness relation~(\ref{complete}).
Actually, it is known that for commuting moduli the latter is easily
violated~\cite{Chu, LP1} unless an additional effort~\cite{TZ} is made.
In fact, the main point for adjusting the noncommutativity of the
translational moduli as in~(\ref{acom}) is to make the completeness
relation work out. To our satisfaction, after (lengthy) computations we can 
indeed confirm the validity of~(\ref{complete}).
In more detail, the matrix $\D\,(\D^\+\D)^{-1}\D^\+$ takes the form
\begin{equation}\label{DD+D-1D+}
\begin{pmatrix}
{\bf1}_2 - \p_n^{-1}{\bf1}_2 
& \p_n^{-1}\frac{\La_1}{r^2_1}x_1^\+ 
& \p_n^{-1}\frac{\La_2}{r^2_2}x_2^\+ &\ldots&
\p_n^{-1}\frac{\La_n}{r^2_n}x_n^\+  \\
x_1\frac{\La_1}{r^2_1}\p_n^{-1} & 
{\bf1}_2 - x_1\frac{\La_1^2}{r^2_1\p_n r^2_1}x_1^\+ & 
-x_1\frac{\La_1}{r^2_1}\p_n^{-1}\frac{\La_2}{r^2_2}x_2^\+
& \ldots& -x_1\frac{\La_1}{r^2_1}\p_n^{-1}\frac{\La_n}{r^2_n}x_n^\+\\
x_2\frac{\La_2}{r^2_2}\p_n^{-1} 
& -x_2\frac{\La_2}{r^2_2}\p_n^{-1}\frac{\La_1}{r^2_1}x_1^\+   
& {\bf1}_2 - x_2\frac{\La_2^2}{r^2_2\p_n r^2_2}x_2^\+  & &\vdots \\
  \vdots &\vdots    & &\ddots & 
-x_{n-1}\frac{\La_{n-1}}{r^2_{n-1}}\p_n^{-1}\frac{\La_n}{r^2_n}x_n^\+
 \\      
x_n\frac{\La_n}{r^2_n}\p_n^{-1}& 
-x_n\frac{\La_n}{r^2_n}\p_n^{-1}\frac{\La_1}{r^2_1}x_1^\+
&\ldots & 
-x_n\frac{\La_n}{r^2_n}\p_n^{-1}\frac{\La_{n-1}}{r^2_{n-1}}x_{n-1}^\+
& {\bf1}_2 - x_n\frac{\La_n^2}{r^2_n\p_n r^2_n}x_n^\+    
\end{pmatrix} \nonumber
\end{equation}
In this computation the relation (\ref{prop_x}) was important.
Likewise, the calculation of $\Psi\Psi^\+$ yields 
\begin{equation}\label{PP+}
\begin{pmatrix}
\p_n^{-1}{\bf1}_2 & -\p_n^{-1}\frac{\La_1}{r^2_1}x_1^\+ 
& -\p_n^{-1}\frac{\La_2}{r^2_2}x_2^\+ &\ldots&
-\p_n^{-1}\frac{\La_n}{r^2_n}x_n^\+  \\
-x_1\frac{\La_1}{r^2_1}\p_n^{-1} & 
x_1\frac{\La_1^2}{r^2_1\p_n r^2_1}x_1^\+ & 
x_1\frac{\La_1}{r^2_1}\p_n^{-1}\frac{\La_2}{r^2_2}x_2^\+
& \ldots& x_1\frac{\La_1}{r^2_1}\p_n^{-1}\frac{\La_n}{r^2_n}x_n^\+\\
-x_2\frac{\La_2}{r^2_2}\p_n^{-1}
& x_2\frac{\La_2}{r^2_2}\p_n^{-1}\frac{\La_1}{r^2_1}x_1^\+   
& x_2\frac{\La_2^2}{r^2_2\p_n r^2_2}x_2^\+  & &\vdots \\
  \vdots &\vdots    & &\ddots & 
x_{n-1}\frac{\La_{n-1}}{r^2_{n-1}}\p_n^{-1}\frac{\La_n}{r^2_n}x_n^\+
 \\      
-x_n\frac{\La_n}{r^2_n}\p_n^{-1}& 
x_n\frac{\La_n}{r^2_n}\p_n^{-1}\frac{\La_1}{r^2_1}x_1^\+
&\ldots & 
x_n\frac{\La_n}{r^2_n}\p_n^{-1}\frac{\La_{n-1}}{r^2_{n-1}}x_{n-1}^\+
& x_n\frac{\La_n^2}{r^2_n\p_n r^2_n}x_n^\+    
\end{pmatrix} \ . \nonumber
\end{equation}
It is not hard to see that the latter two matrices add up to~${\bf1}_{2n+2}$,
as the completness relation~(\ref{complete}) demands.
So, for our choice of~$(\D,\Psi)$ and commutation relations~(\ref{acom}) 
the tensor $F_{\mu\nu}$ is self-dual. 

\medskip

\noindent
{\bf Topological charge.\ } 
In a number of papers (see e.g.~\cite{Chu, Sako})
it was argued that in $\R^4_\th$ the topological charge of 
noncommutative ADHM $n$-instantons for the gauge group U(N) is an 
integer since this is encoded in the dimensions of the ADHM matrices 
even in the noncommutative case. Noncommutativity of translational 
moduli does not alter this conclusion. However, due to the noncommutativity 
of not only $x^\m$ but also $a_i^\m$, our operators are defined on the 
larger Fock space $\Hcal_0\otimes\Hcal_1\otimes\ldots\otimes\Hcal_n$.
That is why the topological charge 
of our solution will be $n$ times the identity operator in 
$\Hcal_1\otimes\ldots\otimes\Hcal_n$. In the commutative 
limit this identity operator becomes unity, and we recover the 
standard 't~Hooft multi-instanton with commutative moduli featuring 
a topological charge equal to~$n$.  

\bigskip

\noindent
{\bf Acknowledgements.\ }
The authors are grateful to A.D.~Popov for fruitful discussions and
for reading the manuscript. 
T.A.I. acknowledges the Heisenberg-Landau Program for partial support and
the Institut f\"ur Theoretische Physik der Universit\"at Hannover for 
its hospitality. This work was partially supported by the Deutsche
Forschungsgemeinschaft (DFG).


\end{document}